\documentclass{aa}
\usepackage[utf8]{inputenc}
\usepackage{microtype}
\usepackage{graphicx}
\usepackage{longtable}
\usepackage{wrapfig}
\usepackage{rotating}
\usepackage[normalem]{ulem}
\usepackage{amsmath}
\usepackage{amssymb}
\usepackage{capt-of}
\usepackage{hyperref}
\usepackage{natbib}
\usepackage{booktabs}
\usepackage[varg]{txfonts}
\usepackage{pbalance}
\usepackage{caption}
\usepackage{siunitx}
\usepackage{booktabs}
\usepackage{longtable}
\usepackage{rotating}
\usepackage[normalem]{ulem}
\usepackage{amsmath}
\usepackage{amssymb}
\usepackage{caption}
\usepackage{subcaption}
\usepackage{siunitx}
\usepackage{orcidlink}
\usepackage[varg]{txfonts}
\bibpunct{(}{)}{;}{a}{}{,} 
\usepackage{booktabs}

\hypersetup{
    colorlinks=true,
    citecolor=blue,
    urlcolor=blue,
    linkcolor=purple,
}

\def\utw{\ensuremath{\smash{\rlap{\lower5pt\hbox{$\sim$}}}}}
\def\udtw{\ensuremath{\smash{\rlap{\lower6pt\hbox{$\approx$}}}}}

\author{
    T. Emil Rivera-Thorsen\orcidlink{0000-0002-9204-3256}\inst{1}\fnmsep\thanks{Corresponding author. \email{trive@astro.su.se}}
\and
    J. Chisholm\orcidlink{0000-0002-0302-2577}\inst{2}
\and
    B. Welch\orcidlink{0000-0003-1815-0114}\inst{3, 4, 5}
\and
    J. R. Rigby\orcidlink{0000-0002-7627-6551}\inst{4}
\and
    T. Hutchison\orcidlink{0000-0001-6251-4988}\inst{4}
\and
    M. Florian\inst{6}
\and
    K. Sharon\orcidlink{0000-0002-7559-0864}\inst{7}
\and
    S. Choe\orcidlink{0000-0003-1343-197X}\inst{1}
\and
    H. Dahle\orcidlink{0000-0003-2200-5606}\inst{8}
\and
    M.~B.~Bayliss\orcidlink{0000-0003-1074-4807}\inst{9}
\and
    G. Khullar\orcidlink{0000-0002-3475-7648}\inst{10}
\and
    M. Gladders\orcidlink{0000-0003-1370-5010}\inst{11, 12}
\and
    M. Hayes\orcidlink{0000-0001-8587-218X}\inst{1}
\and
    A. Adamo\orcidlink{0000-0002-8192-8091}\inst{1}
\and
    M. R. Owens\orcidlink{0000-0002-2862-307X}\inst{9}
\and
    K. Kim\orcidlink{0000-0001-6505-0293}\inst{13}
}

\institute{
    The Oskar Klein Centre, Department of Astronomy, Stockholm University, AlbaNova, 10691 Stockholm, Sweden
    \and
    Department of Astronomy, University of Texas at Austin, 2515 Speedway, Austin, Texas 78712, USA
    \and
    Department of Astronomy, University of Maryland, College Park, MD 20742, USA
    \and
    Observational Cosmology Lab, Code 665, NASA Goddard Space Flight Center, 8800 Greenbelt Rd., Greenbelt, MD 20771, USA
    \and 
    Center for Research and Exploration in Space Science and Technology, NASA/GSFC, Greenbelt, MD 20771
    \and
    Steward Observatory, University of Arizona, 933 North Cherry Avenue, Tucson, AZ 85721, USA
    \and 
    Department of Astronomy, University of Michigan, 1085 S. University Ave, Ann Arbor, MI 48109, USA
    \and
    Institute of Theoretical Astrophysics, University of Oslo, P.O. Box 1029, Blindern, NO-0315 Oslo, Norway
    \and
    Department of Physics, University of Cincinnati, Cincinnati, OH 45221, USA
    \and
    Department of Physics and Astronomy and PITT PACC, University of Pittsburgh, Pittsburgh, PA 15260, USA
    \and
    Department of Astronomy and Astrophysics, University of Chicago, 5640 South Ellis Avenue, Chicago, IL 60637, USA
    \and
     Kavli Institute for Cosmological Physics, University of Chicago, Chicago, IL 60637, USA
     \and 
     IPAC, California Institute of Technology, 1200 E. California Blvd., Pasadena CA, 91125, USA
}

\titlerunning{Wolf-Rayet stars in the Sunburst Arc}
\authorrunning{T. E. Rivera-Thorsen et al.}
\date{Recievied April 13., 2024; Accepted August 5., 2024.}
\title{The Sunburst Arc with JWST: I. Detection of Wolf-Rayet stars injecting nitrogen into a low-metallicity, $z=2.37$ proto-globular cluster leaking ionizing photons}
\hypersetup{
 pdfauthor={T. Emil Rivera-Thorsen},
 pdftitle={Wolf-Rayet stars in the Sunburst Arc},
 pdfkeywords={},
 pdfsubject={},
 pdflang={English}}

\abstract
{
We report the detection of a population of Wolf-Rayet (WR) stars in the Sunburst Arc, a strongly gravitationally lensed galaxy at  redshift $z=2.37$.  
As the brightest known lensed galaxy, the Sunburst Arc has become an important cosmic laboratory for studying star and cluster formation, Lyman $\alpha$ (Ly$\alpha$) radiative transfer, and Lyman Continuum (LyC) escape.

Here, we present the first results of JWST/NIRSpec IFU observations of the Sunburst Arc, focusing on a stacked spectrum of the 12-fold imaged LyC-emitting (Sunburst LCE) cluster.
In agreement with previous studies, we find that the Sunburst LCE cluster is a very massive, compact star cluster with $M_{\text{dyn}} = (9\pm1) \times 10^{6}
M_{\odot}$. 
Our age estimate of 4.2--4.5 Myr is much larger than the crossing time of $t_{\text{cross}} = 183 \pm 9 $ kyr, indicating  that the cluster is dynamically evolved and consistent with being gravitationally bound. We find a
significant nitrogen enhancement of the low ionization state interstellar medium (ISM), with \(\log(N/O) =
-0.74 \pm 0.09\), which is \(\approx 0.8\) dex above typical values for H\,\textsc{ii} regions of
similar metallicity in the local Universe. We find broad stellar emission
complexes around He\,\textsc{ii}$\lambda 4686$ and C\,\textsc{iv}$\lambda 5808$ with associated
nitrogen emission --- this is the first time WR signatures have been directly observed at redshifts above $\sim 0.5$. The strength of the WR signatures cannot be reproduced by stellar population models that only include single-star evolution. While models with binary evolution better match the WR features, they still struggle to reproduce the nitrogen-enhanced WR features.  JWST reveals the Sunburst LCE to be a highly ionized proto-globular cluster with low oxygen abundance and extreme nitrogen enhancement that hosts a population of Wolf-Rayet stars, likely including a previously suggested population of Very Massive Stars (VMSs), which together are rapidly enriching the surrounding medium. }
 
\begin{document}
\maketitle

\section{Introduction}
\label{sec:org9e5059e}
Massive stars dominate the mechanical feedback,  ionizing photon production, and chemical enrichment of young stellar populations. Such stars spend about $\sim10\%$ of their lives in an evolutionary phase where they are classified as Wolf-Rayet (WR) stars \citep{meynet05}.  Classical WR stars are massive, evolved, very hot stars that have lost their hydrogen envelopes \citep[e.g., ][]{crowther07}.  
Typically, there are two formation channels to shed the outer hydrogen layers of WR stars: through extreme stellar wind mass loss \citep{meynet05} or through close binary interactions. Metal-line absorption drives gas off the surface of massive stars;  the mass-loss rate is extremely sensitive to the metallicity of the star. 
Thus, though the WR phase is a fleeting part of the short life of a massive star, collectively WR stars can significantly contribute to or even dominate the winds, ionizing photons, and nucleosynthetic production of star-forming populations.

Individual WR stars have been identified out to distances of $\lesssim 5$\,Mpc \citep[e.g.,][]{schootemeijer2018,dellabruna2021,dellabruna2022b}.  WR signatures have been identified in the spectra of nearby galaxies \citep[e.g.,][]{vacca1992,gomezgonzalez2021}, 
and as far out as redshift $z = 0.45$ \citep[e.g.,][]{menacho2021,yuan2022}. WR stars are most common at higher metallicities, but WR signatures have been observed in low-metallicity galaxies \citep[e.g.,][]{amorin2012,kehrig2013} and even in I Zwicky 18, the most metal-poor known galaxy \citep{guseva2000}, demonstrating that such stripped stars are readily produced, despite the dependence of line--driven mass loss on metallicity.  Binary stellar evolution paths may be particularly important at low metallicity.

Stacked rest-frame ultraviolet (UV) spectra of $z\sim3$ galaxies have revealed broad stellar wind features that may be due to classical WR stars \citep{shapley2003,rigby2018b}, but could also be explained by the presence of Very Massive Stars \citep[VMSs; e.g.,][]{wofford2014,crowther2016,smith2023,upadhyaya2024}. VMS are conventionally defined as main-sequence stars with $M \ge 100 M_{\odot}$. 
Unlike lower mass main-sequence stars, VMS have strong stellar winds which make them appear spectroscopically similar to WR stars. They have not yet shed their hydrogen envelopes, so while they do display strong wind features like classical WR stars, they do not share the lacking hydrogen lines and richness in metal emission features of the latter. 
While the broad wind features in the rest-frame UV are almost identical for VMSs and WR stars, similar wind features exist in the rest-frame optical, which can help discriminate between the two star types \citep{martins2023}.
However, while \citeauthor{martins2023} present a classification scheme to distinguish between the two types of stars, \cite{vink2023} argues that VMSs are the same objects as the WNh type WR stars, and later evolve into classical nitrogen rich WR stars of type WN. 

What has been missing, until now, is rest-frame optical spectra of the distant universe with sufficient sensitivity and dynamic range to reveal the faint, subtle signs of WR stars within their stellar populations. 

In this paper, we present rest-frame optical James Webb Space Telescope (JWST)/NIRSpec Integral Field Unit (IFU) spectra of a bright, massive star cluster and surrounding \ion{H}{ii} region within the Sunburst Arc \citep{dahle2016}, a strongly star forming galaxy at \(z = 2.37\) which is gravitationally lensed by a foreground ($z=0.44$) galaxy cluster \citep{sharon2022,pignataro2021,diego2022}.  The region targeted is known to be leaking ionizing photons \citep{riverathorsen2019}.  By stacking the spectrum of multiple lensed images of this region, we obtain a spectrum that is directly comparable, in signal-to-noise, wavelength range, and spectral resolution, to the spectra of \ion{H}{ii} regions in the best--studied nearby galaxies. 
The galaxy morphology is dominated by a number of emission clumps, of which especially one is very bright \citep{riverathorsen2019,vanzella2022b,pignataro2021}. Due to the large shear magnification, this bright clump is resolved down to pc scales, revealing a compact star cluster with a stellar population age of $\sim 3.6$ Myr \citep{chisholm2019}, and steep stellar UV slope $\beta \approx -3$ \citep{kim2023}.
\cite{chisholm2019} found broad C\,\textsc{iv}$\lambda$ 1548 and H\,\textsc{ii}$\lambda$ 1640 stellar wind lines in rest-frame UV Magellan/MagE spectra, but it is very challenging to discriminate between the two kinds of stars from these lines alone. 
\cite{pascale2023,mestric2023} attribute the rest-UV stellar features in the Sunburst LCE to VMSs. \cite{pascale2023} find very strong nitrogen enhancement in hot, high-density gas condensations that they argue must be located outside, but in near proximity to, the cluster. Such very strong and highly localized nitrogen enrichment has also been observed at high redshifts since the launch of JWST, and is often, but not always, considered consistent with being injected by VMSs and WR stars \citep[e.g.,][]{kobayashi2024,ji2024}.

\begin{figure*}
\centering
\includegraphics[width=0.95\textwidth]{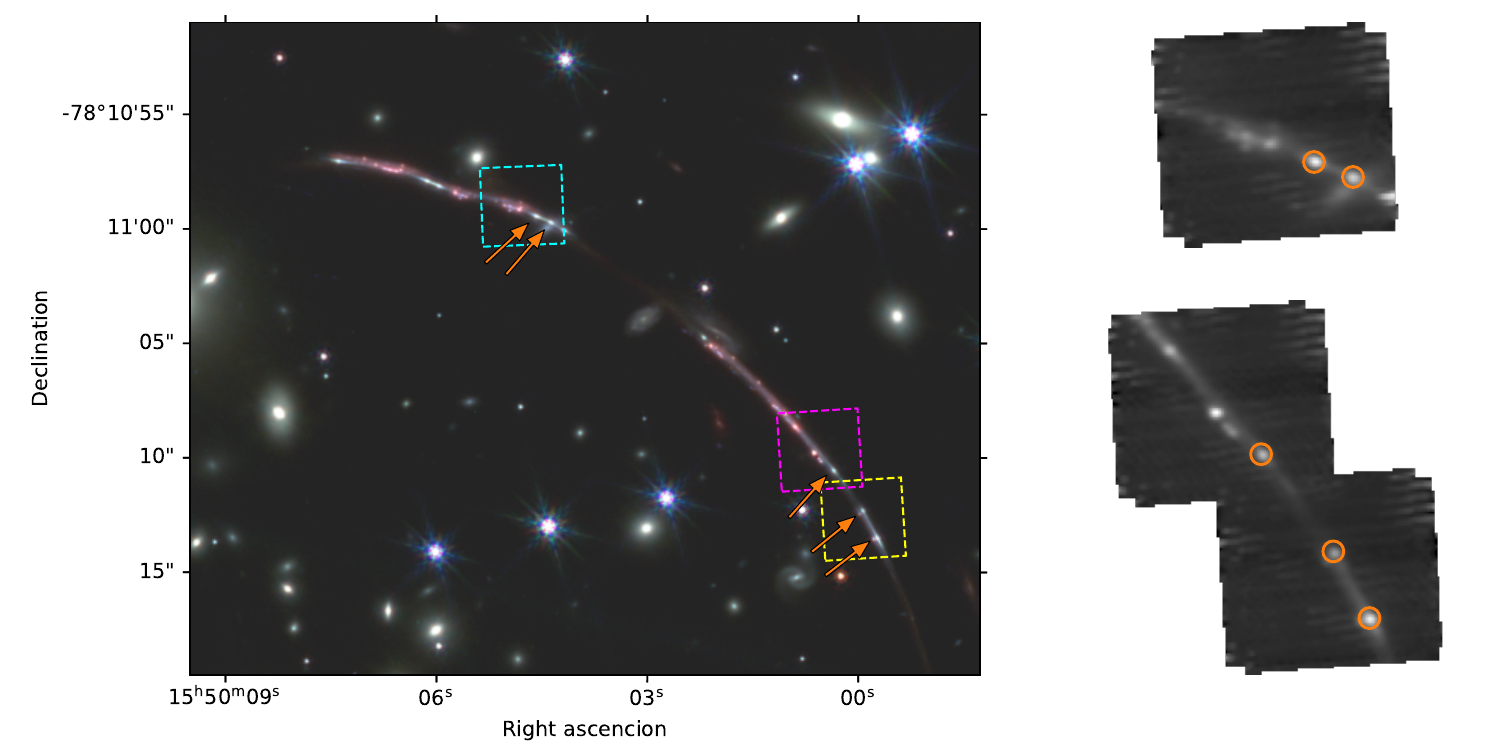}
\caption{\label{fig:overview}Overview of the arc. \textbf{Left:} NIRCam RGB composite of the N and NW arc segments, with R, G, B being \textbf{F444W, F200W, and F115W}, respectively, together covering the rest-frame wavelength range 3400--13000 Å. Cyan, magenta, and yellow dashed overlays show the approximate footprints of the three NIRSpec pointings 1, 2 and 3, respectively. Orange arrows show the images of the gravitationally lensed LCE cluster which were included in the stacked spectrum. \textbf{Right:} NIRSpec IFU continuum images of pointings 1 (\textbf{top right}) and combined pointings 2+3 (\textbf{bottom right}); created from a median stacking along the spectral axis of the F100L/G140H cubes. Orange circles mark the images of the LCE cluster included in the stacked spectrum.}
\end{figure*}

\section{Observations and data reduction}
\label{sec:orga1d90cc}

The target was observed with JWST/NIRCam imaging and JWST/NIRSpec integral field spectroscopy in JWST Cycle 1, as part of program GO-2555 (PI: Rivera-Thorsen) over the
time period of April 4, 2023 to April 10, 2023. 
The entire field was observed in each of the NIRCam filters F115W, F150W, F200W, F277W, F356W, and
F444W. Four pointings were observed with NIRSpec: three on-target covering different sections of the  Sunburst Arc, and one
off-target for background correction (\autoref{fig:overview}). The target was observed in each of the
settings F100L/G140H and F170L/G235H, covering a combined rest-frame wavelength range of 2900 Å \(\le \lambda_0 \le\) 9700 Å; however, for the sake of this analysis, only the wavelength range 3500 Å \(\lesssim \lambda_0 \lesssim\) 8000 Å was used. 
The spectral resolving power within this wavelength interval ranges from \(R \approx 2250\; (\Delta v \approx 133 \text{km s}^{-1})\) to \(R \approx 3820\; (\Delta v \approx 78\, \text{km s}^{-1})\) for g140h; and from \(R \approx 1690\; (\Delta v \approx 160\, \text{km s}^{-1})\), to \(R\approx 3140\; (\Delta v \approx 80\, \text{km s}^{-1})\) for g235h. The PSF of NIRSpec is not yet well characterized; but recent treatment by \cite[see their Fig.~7]{deugenio2023arXiv} suggests the PSF is stable in size up to \(\lambda \approx 3 \mu\)m, with a minor-axis Full Width at Half Maximum \(\text{FWHM} \approx 0\farcs 1\), and a major-axis \(\text{FWHM} \approx 1\farcs 4\). For comparison, the native spaxel size of the IFU is \(0\farcs 1\).

We refer the reader to the main companion paper (Rivera-Thorsen et al., in
prep.) for a thorough explanation of the data reduction and calibration steps, but outline the most important steps here.
We reduced the NIRSpec IFU observations following the methods described in \cite{rigby23_overview}, using the \textsc{Templates} NIRSpec data reduction notebooks \citep{templates_data_notebooks}. For the reduction, we used version \texttt{1.11.4} of the JWST pipeline \citep{bushouse23_pipeine1p11p4}, and the calibration reference files from context \texttt{pmap\_1105}. After completing the pipeline reduction, we removed remaining outliers from the data cubes using the \texttt{baryon-sweep} software \citep{hutchison23_sigmaclip, baryonsweep}.

\section{Emission line modeling}
\label{sec:org3bbece7}
We extracted a stacked spectrum from five of the six lensed images
of the LCE cluster covered by the IFU (\autoref{fig:overview}).In the nomenclature of
\cite{riverathorsen2019,sharon2022}, they are images 4, 5, 8, 9, and 10. Image 6 was
partially covered by the IFU, but since the data quality was negatively affected
by its position right at the edge of the detector, we omitted this image from our analysis.
We extracted a spectrum from an area of $5\times5$ spaxels covering each image of the
LCE cluster. For each spectrum, the aperture spaxels were continuum-weighted using the median value along the spectral
axis before extraction. Each extracted spectrum was then normalized by the median flux value in the wavelength bins in the wavelength overlap interval between the two disperser settings.
Since the spectra are differently magnified, we have made no attempt to preserve absolute fluxes, and all results in this work are based on flux ratios.
The extracted spectrum displays extremely high signal-to-noise ratio and
sensitivity. We detect stellar continuum at a comfortable signal-to-noise ratio across the entire wavelength range, and have identified 59 emission lines in this spectrum, including
Balmer- and Paschen lines from up till level 20, as well as auroral emission
lines from multiple species and ionization levels. A full analysis of abundances
and ionization zones based on direct $T_e$ methods are presented in a companion paper \citep{welch2024}.

\begin{figure*}
\centering
\includegraphics[width=\textwidth]{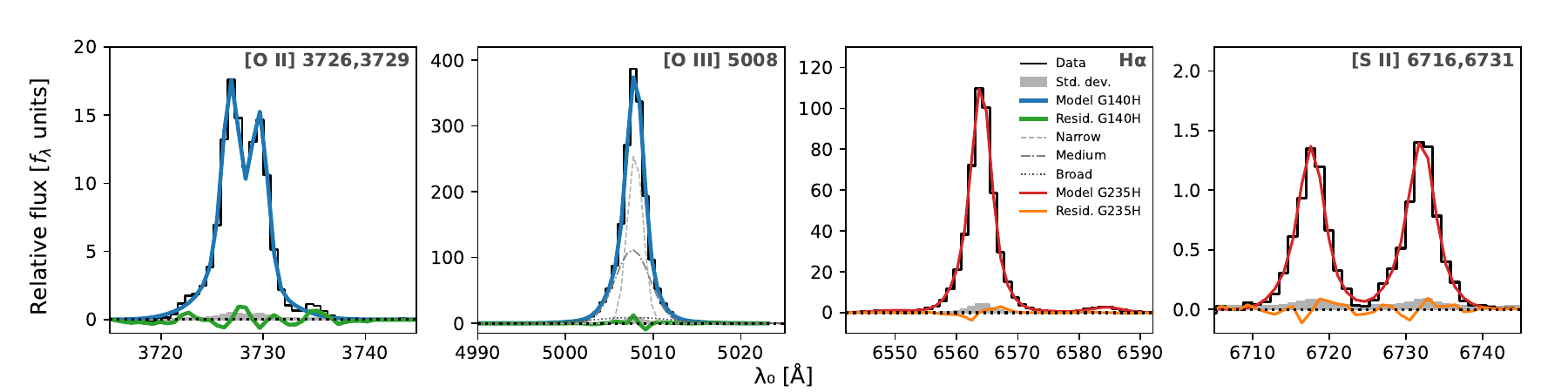}
\caption{\label{fig:fit}Best-fit model of selected emission lines, shifted to rest-frame wavelengths. Here are shown the two strongest lines dominating the kinematics, as well as the two \(n_{e}\)-sensitive doublets, [O\,\textsc{ii}] 3727,3729; and [S II] 6716,6731.}
\end{figure*}

All subsequent line ratio based computations are based on the full flux of each line; but to account for all flux in each line as accurately as possible, we have modeled the line emission by three kinematic components, each modeled
with shared FWHM and redshift, with flux in different lines left as a free
parameter, such that for \(N\) emission lines with \(M\) kinematic components each
included in the fit, we have a number of free parameters \(n_{\text{pars}} = M(N +
2)\). In some faint lines, the broadest components have fallen below the noise level and thus only added noise to the fit, in which case they were removed.
We originally included only 2 components; one narrow and one broad; but it
was necessary to add a third, broad component in order to get a reasonable fit
of the brightest lines, as especially H\(\alpha\) contains a non-negligible broad
component of FWHM \(\approx\) 800 km s$^{-1}$ which likely originates from outflows driven by stellar winds from massive evolved stars in the cluster. 
We corrected the inferred line widths for
instrument broadening using the official JWST/NIRSpec calibration files
available for download at J-Docs\footnote{\url{https://jwst-docs.stsci.edu/jwst-near-infrared-spectrograph/nirspec-instrumentation/nirspec-dispersers-and-filters}} (these are the versions delivered June 2016 and are at the time of writing the most current). 
We interpolated the tabulated values of
\(R\), and obtained the value corresponding to the observed wavelength of each
line. We assumed Gaussianity and corrected for instrument broadening by adding
in quadrature the velocity dispersion and the instrument broadening in the model of
each line to obtain the intrinsic line width.

\autoref{fig:fit} shows a comparison of the observed data and the best-fit
model for a few selected lines: [O\,\textsc{iii}] 5008 and H\(\alpha\), which are among the lines
dominating the kinematics of the fit; and the two \(n_{e}\)-sensitive doublets
[O\,\textsc{ii}] 3727,3729 and [S II] 6717,6731. Especially the [O\,\textsc{ii}] doublet, which is
blended at the resolving power of NIRSpec, can often have its inferred flux ratio be sensitive to an ill-fitting kinematic model. However, as is readily seen in the figure, the kinematics of the
strongest lines fits well with both doublets, and we adopt the line ratios
derived from these fits with good confidence.

To resolve the narrow emission to the best possible extent, we also performed a
fit to the [O\,\textsc{iii}] \(\lambda \lambda\)4960,5008 in F100L/G140H, where the resolving power is
highest at \(R \approx 3300\). This fit yielded FWHM = 83 km s$^{-1}$, or \(\sigma_{\text{LOS}} \simeq 35.3 \pm 2.1\)
km s$^{-1}$, fully consistent with the value obtained by \citeauthor{vanzella2022b} using
VLT/X-shooter.

\begin{table}[htbp]
\caption{\label{tab:kinem}best-fit kinematic properties of the three components.}
\centering
\begin{tabular}{lrcc}
\toprule
Comp. & $z$ & $\Delta v\,$\tablefootmark{a} & FWHM\tablefootmark{a} \\[0pt]
\midrule
Narrow\tablefootmark{b} & $2.371062 \pm 6 \times 10^{-6}$ & --- & $91 \pm  7$\\
Medium & $2.37093 \pm 2 \times 10^{-5}$ & $-11.7 \pm 1.9$ & $277 \pm 6$\\
Broad & $2.3709 \pm 2 \times 10^{-4}$ & $-14.4\pm17.8$& $708 \pm 22$\\
\bottomrule
\end{tabular}
\tablefoot{\\
\tablefoottext{a}{Given in km s$^{-1}$ relative to the narrow component.}\\
\tablefoottext{b}{The best-fit narrow component FWHM from [O\,\textsc{iii}] in G140H is 83 \textpm{} 4.9 km s$^{-1}$}
}
\end{table}

\begin{table}[htbp]
\caption{\label{tab:lineflux}Measured line fluxes relative to H\(\beta\).}
\centering
\begin{tabular}{rcc}
\toprule
Line & Relative flux [\%] & Uncertainty [\%] \\
\midrule
$[\text{O \scshape{ii}}]$ 3727 & $30.66$ & $5.31$\\
$[\text{O \scshape{ii}}]$ 3729 & 22.87 & 5.91\\
$[\text{O \scshape{iii}}]$ 4363 & 10.94 & 6.07\\
 H$\beta$ 4861 & 100 & --\\
$[\text{O \scshape{iii}}]$ 4960 & 237.58 & 19.15\\
$[\text{O \scshape{iii}}]$ 5008 & 707.32 & 56.67\\
 H$\alpha$ 6562 & 334.17 & 27.59\\
$[\text{N \scshape{ii}}]$ 6548 & 4.81 & 0.79\\
$[\text{N \scshape{ii}}]$ 6584 & 11.07 & 1.18\\
$[\text{S \scshape{ii}}]$ 6717 & 4.20 & 0.52\\
$[\text{S \scshape{ii}}]$ 6731 & 4.47 & 0.53\\
\bottomrule
\end{tabular}
\end{table}

Tables \ref{tab:kinem} and \ref{tab:lineflux} show the results of the modeling; the
former shows the centroid and line widths of the three kinematic components,
while the latter shows the total measured flux of the measured lines relative to
the measured flux in H\(\beta\). 

\section{LCE Cluster properties}
\subsection{Detection of Wolf-Rayet Stars}
\label{sec:org019537e}

\begin{figure*}
\centering
\includegraphics[width=.9\linewidth]{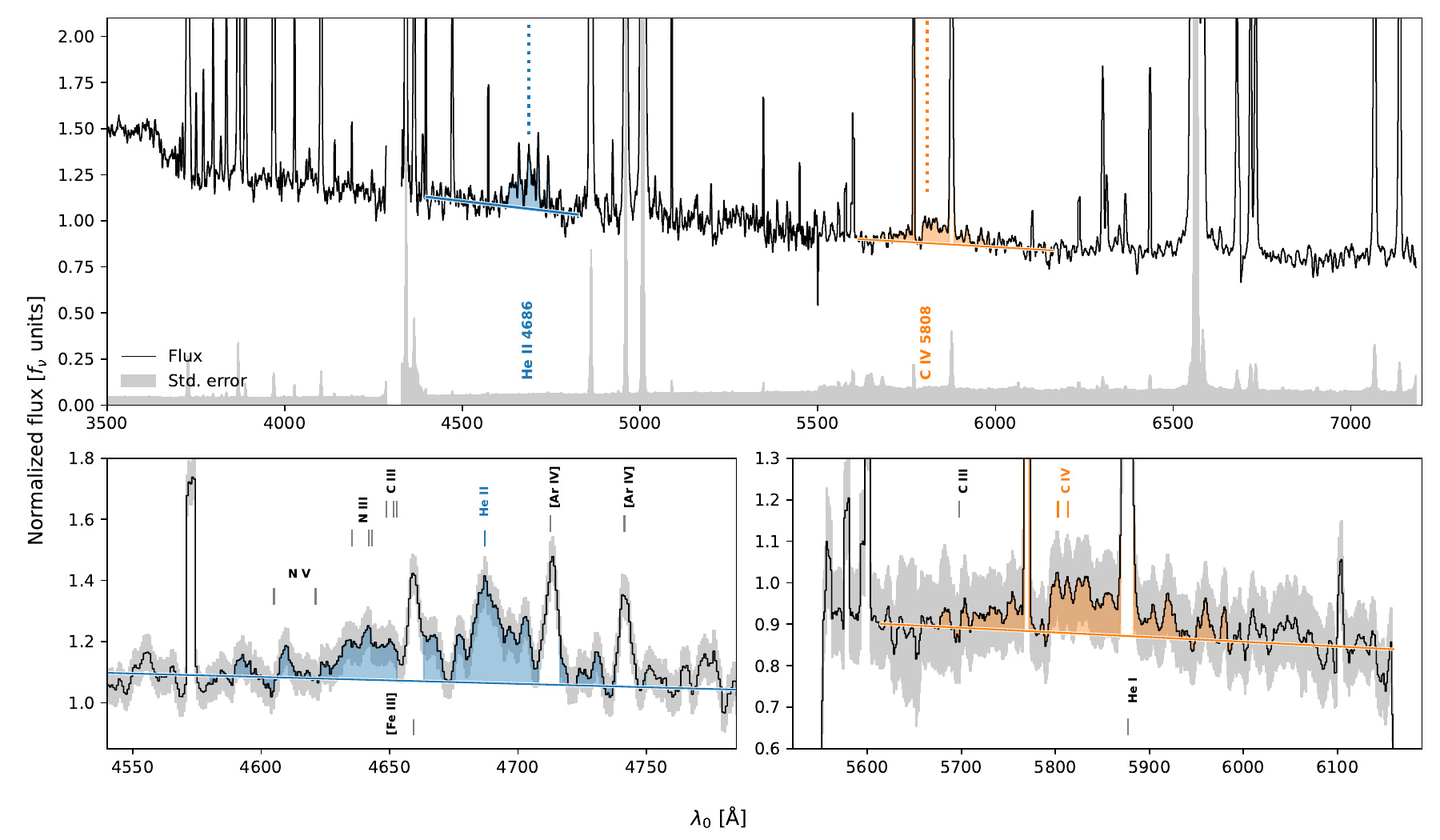}
\caption{\label{fig:wrfig}Extracted and stacked spectrum of the Sunburst LCE, zoomed into the continuum to clearly show the blue and orange WR bumps. Blueward of 5500 Å, only the G140H spectrum is shown; redward, only the G235H spectrum is shown. Almost all the spikes are emission lines. The best-fit continuum at each bump is shown in its corresponding color. Lower left: Detailed view of the blue WR bump; the gray shading shows the $\pm 1 \sigma$ errors. A number of emission line centroids are marked. 
Lower right: A similar detailed view of the orange WR bump, centered on the C\,\textsc{iv} \(\lambda \lambda\) 5801,5812  Å feature. On its red side is a He I \(\lambda\) 5875 emission line; on the blue side an instrument artifact. The blue and orange shading of the bumps is mainly provided to guide the eye.}
\end{figure*}

Previous ground-based, rest-frame UV spectroscopic observations with Magellan and the VLT have shown broad stellar emission in He\,\textsc{ii} \(\lambda\) 1640 and C\,\textsc{iv} \(\lambda\) 1548 \citep{chisholm2019,mestric2023}. These lines are typically attributed
to Wolf-Rayet stars, but can also be attributed to VMSs; usually
defined as main sequence stars of \(M_{\star} \ge 100\, M_{\odot}\). The emission has been attributed
to VMSs by, e.g., \citet{mestric2023} and \cite{pascale2023} from secondary
evidence; but it is very difficult to distinguish these two scenarios from the
rest-frame UV emission alone. 
This degeneracy can, however, be broken with rest-frame optical spectral features.
The classical Wolf-Rayet ``bumps'' at $\sim 4686$ Å (the ``blue bump'') and $\sim 5808$ Å (the
``yellow'', ``orange'' or sometimes ``red bump''; here we adopt the term ``orange'') consist of complexes of metal lines centered around He\,\textsc{ii} \(\lambda\) 4686 (blue) and C\,\textsc{iv}
\(\lambda\) 5808 (orange), which are present at varying strengths depending on the metal
abundances, ionization/temperature, and other properties of the stars from which
they originate. This allows for a distinction between VMS and WR stars, as
outlined in the classification scheme by \citet{martins2023}. According to this scheme, VMS give rise to a blue bump dominated by stellar \ion{He}{ii}, and the orange bump, if present at all, is weak and consisting of a simple \ion{C}{iv} doublet; the dense complex of surrounding metal lines is largely absent.

\autoref{fig:wrfig} shows the stacked spectrum of the Sunburst LCE cluster in the
relevant wavelength intervals. The spectrum shows a clear, broad blue bump, and a somewhat weaker broad orange bump. The two features are
shown in close-up in the lower panels. Two forbidden, nebular Argon lines [Ar\,\textsc{iv}]\(\lambda \lambda\)~4711,4740 coincide with the red side of the blue bump, but are not part
of this line complex; neither is the narrow, nebular He\,\textsc{i} \(\lambda\) 5875 line.

To measure line width and flux of the WR \ion{He}{ii} 4686 feature, we have fit it a single Gaussian profile. We have enforced the continuum level shown
in the upper panel, and included the wavelength interval between [Fe\,\textsc{iii}] and
[Ar\,\textsc{iv}] only. The fit yields a velocity \(\text{FWHM} = 1370 \pm 250 \text{km s}^{-1}\)
and restframe \(\text{EW} = 4.1 \pm 0.4\) \AA{}. Following the methodology of \citeauthor{martins2023}, the widths and relative strengths of the bumps, and in particular the prevalence of the permitted N\,\textsc{iii} features at $\lambda\lambda$ 4620,4640, all place these features firmly in the Wolf-Rayet category. However, since what separates the two stellar types in this scheme is mainly the absence of certain features in VMS compared to WR stars, having found strong spectroscopic evidence for WR stars does not rule out a --- potentially significant --- contribution from VMS; see Sect.~\ref{sec:wrprops}.

\label{sec:org1c100d4}
\subsection{Dust attenuation}
\label{sec:org0578072}
To model dust attenuation, we have adopted a standard starburst attenuation law
\citep{calzetti2000} with a standard value of \(R_{V} \equiv A_{V} / E(B-V) = 4.05\). 
We have computed E(B-V) following \citet{calzetti2000}. 
Adopting a standard intrinsic \(H\alpha/H\beta = 2.86\) yields a value of \(E(B-V) = 0.11 \pm
0.19\),
which we have adopted to deredden line fluxes.

\subsection{Dynamical mass}
\label{sec:org4722f56}
Following \citet{rhoads2014,vanzella2022b}; we find the dynamical mass using
the approximation \(M_{\text{Dyn}} \approx 4\sigma_{\text{LOS}}^{2}R_{\text{eff}}/G\); with
\(R_{\text{eff}}\) being the effective radius of the LCE cluster. Since we do not have a measurement of the stellar velocity dispersion, we have adopted the
line width of the narrow component of the strong emission lines \(H\alpha, H\beta\), and
the [O\,\textsc{iii}] doublet at 4960,5008 \AA{} as a proxy for \(\sigma_{\text{LOS}} = 38.6 \pm 2.9\,\text{km s}^{-1}\). 
Adopting the value \(R_{\text{eff}} = 7.8\) pc \citep{mestric2023}, we find \(M_{\text{dyn}} = (9.0 \pm 1.1) \times 10^{6} M_{\odot}\). 

\subsection{Oxygen abundance}
\label{sec:org1c4b411}
We have estimated the oxygen abundance in the gas based on the direct T\textsubscript{e} method,
following the recipe of \cite{perezmontero2017}, using the O\(^{+ }\) and
O\(^{++}\) based abundances together; we have not observed any nebular He II and
thus believe the contribution of higher ionization stages is negligible.

We have first found electron temperature and density using \texttt{PYNEB}
\citep{pyneb2015} and its \texttt{getCrossTemDen} method, which yielded
\(T_e = 14\pm3\, \rm{kK}\) and \(n_{e}(\text{O}^{+}) = 1440^{+1360}_{-870}\, \text{cm}^{-3}\). Using [S\,{\sc ii}]\(\lambda\lambda\,6716,6731\) as an
\(n_{e}\) diagnostic instead yields an identical value of \(T_{e}\), and an electron density of
\(n_{e}(\text{S}^{+ }) = 1040^{+620}_{-490}\, \text{cm}^{-3}\).

Following eqs. 38 and 40 in \citet{perezmontero2017}, we find:

$$12 + \log\left(\frac{\text{O}^{+ }}{\text{H}^{+ }}\right) = 6.65^{+0.39}_{-0.27};  \quad 
12 + \log\left(\frac{\text{O}^{2+ }}{\text{H}^{+ }}\right) = 7.95^{+0.32}_{-0.22} $$

\noindent and from the approximation O/H \(= (\text{O}^{+} + \text{O}^{2+
})/\text{H}^{+}\) \citep{izotov2006,perezmontero2017}, we get
\(12 + \log(\text{O/H}) = 7.97^{+0.33}_{-0.21}\).

Adopting the solar value of \(12+\log(\text{O/H}) = 8.69\) from \citet{asplund2009},
this yields an oxygen abundance for the sunburst LCE cluster gas of about \(0.19^{+0.20}_{-0.07}\; Z_{\odot}\).
\citet{chisholm2019} found a stellar metallicity of \(Z_{\star} \approx 0.3\; Z_{\odot}\)
and \citet{pascale2023} found a nebular metallicity of \(Z_{\text{neb}} \approx 0.26\, Z_{\odot}\) --- both higher than the nebular abundance found in this work, but fully consistent within the error bars.
From the derived temperature and density, we have also derived the pressure, and found $P_{\text{ion}}/k_B = (3.6^{+2.6}_{-1.6}) \times 10^7 \rm{K\;cm}^{-3}$. This is consistent with values in the centers of local Universe starburst galaxies \citep{dellabruna2021,dellabruna2022b}. 

\subsection{Nitrogen enhancement}
\label{sec:org2dfda50}

Despite having low metallicity, the Sunburst LCE cluster is highly nitrogen
enriched. \citet{pascale2023} report \(\log(\text{N/O}) = -0.23_{-0.11}^{+0.08}\) based on
rest-frame UV N\,\textsc{iii}] emission from VLT/X-shooter spectroscopy.

Here, we have computed \(\log(\text{N/O})\) from [O\,\textsc{ii}]\(\lambda \lambda\)~3727,3729 and
[N\,\textsc{ii}]\(\lambda\)~6584 following \citet{perezmontero2017}. This ion abundance ratio is
expected to follow the true element abundance closely because of the strong 
similarity in ionization potentials between the two species. Still following
\citet{perezmontero2017}, we computed  

$$t([\text{N\,\textsc{ii}}]) \equiv 10^{-4} T_{e}(\text{N}^{+ }) =
\frac{1.85}{[t(\text{[O~\scshape{iii}]}) - 0.72]}$$

\noindent and adopted this as the low-excitation temperature in the nomenclature of those
authors\footnote{Adopting the [O\,\textsc{iii}]\(\lambda\) 4363 -based value of \(T_e\) only changed the end result marginally.}. We derive an [N\,\textsc{ii}] -based nitrogen abundance for the Sunburst
LCE of \(\log(\text{N/O}) = -0.74 \pm 0.09\). 
This value is \(\sim 0.7\) \citep{berg2012} to 0.8 \citep{topping2024} dex above
the typical value for local Universe H\,\textsc{ii} regions with the same oxygen abundance
12+log(O/H) $\approx$ 7.95; but still half a dex below the even higher N\,\textsc{iii}] based value found
by \citet{pascale2023}. We find that this corroborates the hypothesis put
forward by \citeauthor{pascale2023}, that the majority of the nitrogen enhancement in the Sunburst LCE 
is found in dense, highly ionized clouds in close vicinity of the LCE cluster, the
most massive members of which are currently the main drivers of this nitrogen
enhancement.A similar scenario has also been suggested recently for high-redshift
galaxies in general by \citet{ji2024}, who also suggest that the surrounding,
low-density gas has a more normal chemical evolution.

Recent observations of galaxies at $z > 10$ have found extreme N/O enhancements only a few 100~Myr after the Big Bang that has been compared to the Sunburst Arc \citep{cameron23, senchyna23, marqueschaves2024, castellano2024}. Nitrogen is often thought to be produced by long-lived evolved lower mass stars, and these extreme N/O observations in the early universe have questioned the origin of early nitrogen production. The observations of WR stars within the Sunburst Arc point to a plausible causal origin of the nitrogen: extreme nitrogen production from classical WR and WNh stars \citep{berg2012, kobayashi2024}. This suggests that short-lived evolved massive stars may play a crucial role in the build-up of metals within the first few 100~Myr of cosmic history and might hold a key to inferring the properties of the first galaxies formed in the universe. 

In this work, we find that while the nitrogen enhancement in the low-density medium is far lower than in the denser medium, it is still quite strongly enhanced. 

This also challenges one part of the scenario proposed
by \citeauthor{pascale2023}: that the dense, ionized medium is located outside and
at a distance from the LCE cluster. Such a scenario would see a higher Nitrogen
abundance in the neutral, low-density medium compared to the dense medium than
we observe here. Rather, this gas might be dense condensations within the
LCE cluster itself, in close proximity to the most massive stars.

\subsection{WR population properties}
\label{sec:wrprops}
\begin{figure*}
\centering
\includegraphics[width=.99\textwidth]{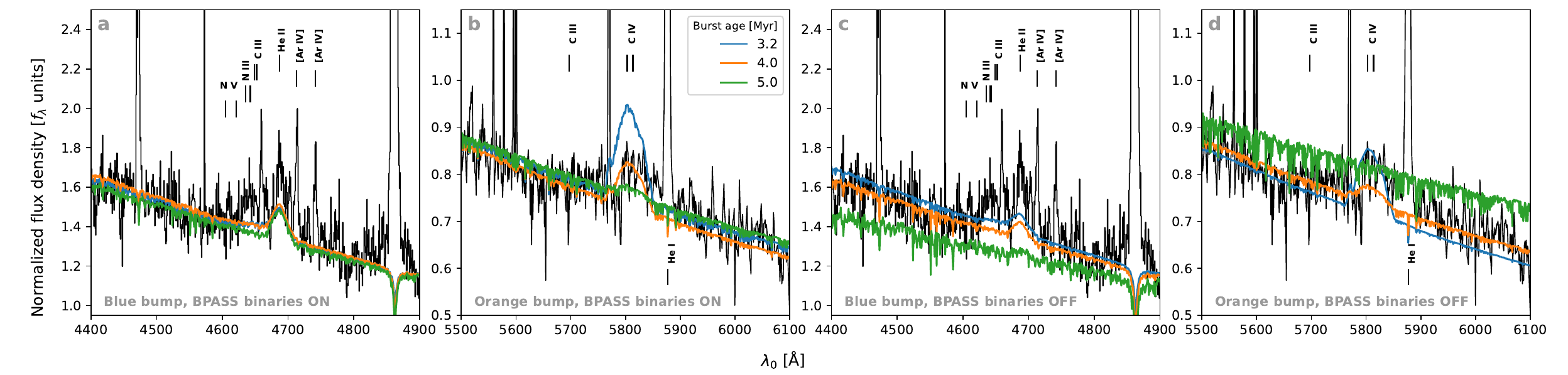}
\caption{\label{fig:bpass}The blue (\textbf{a}, \textbf{c}) and orange (\textbf{b}, \textbf{d}) WR bumps in the Sunburst LCE, this time shown in normalized \(f_{\lambda}\) units. Along with the observed data is shown a selection of the fiducial \texttt{BPASS} \citep{eldridge2017,stanwayeldridge2018,hoki2020} synthetic spectra corresponding to a metallicity of 19\% \(Z_{\odot}\) as found from the direct method, and the range of ages at which a considerable WR population can be present. The models in panels \textbf{a} and \textbf{b} have binary evolution enabled, in panels \textbf{c} and \textbf{d} they are disabled.}
\end{figure*}

To investigate the properties of the WR population, we have compared the WR bumps
to synthetic spectra produced by the \texttt{BPASS} code
\citep{eldridge2017,stanwayeldridge2018,hoki2020}, see \autoref{fig:bpass}. 
These spectra were taken
from the ``fiducial'' suite of models, produced using a Kroupa IMF with slope of $-2.35$, minimum and maximum stellar masses of 0.5, 300 \(M_{\odot}\), and a metallicity of $Z = 0.04 \approx 20\%\,Z_{\odot}$ \citep{hoki2020}. For each time step, we show the synthetic spectrum with binary evolution tracks enabled (leftmost panels), and disabled (rightmost panels). 

Following the method outlined by
\citet{delvalleespinosa2023}, we have visually compared \texttt{BPASS} age steps in the range \(6.3 \le \log(\text{age [yrs]}) \le 6.8\), during which WR stars are prevalent, to the observed
spectrum.  In particular, we have looked for the age step at which the relative
strengths of the two emission features is best emulated by the model spectra.
Using the relative heights of the bumps as a proxy for their relative fluxes, we
found that the observed spectra had a peak ratio between the bumps of
\(I(4686)/I(5808)_{\text{obs}} \approx 2.35\). The model spectrum yielded
\(I(4686)/I(5808) \approx 1.79\) at 4 Myr burst age, and \(I(4686)/I(5808) \approx 8.24\) at 5
Myr burst age. 
The observed spectrum is sandwiched between these two ages, but
closer to 4 Myr than 5 Myr of age. For comparison,
\citet{chisholm2019} found a light-weighted cluster age of 3.6 Myr from
fitting a similar class of \texttt{BPASS} models to stacked, rest-frame UV slit spectra of the
Sunburst LCE. The difference of $\sim 0.5$ Myr is not significant, as on these short timescales the  assumption of an instantaneous starburst likely breaks down.

The best fitting \texttt{BPASS} model population corresponding to 19\% solar metallicity and a population age of 4 Myr contains (adjusted for mass) $\sim 900$ WR stars, with the fractions of WNh, WN, and WC stars being 68\%, 19\%, and 13\%, respectively. 
Assuming that VMS are indeed WNh stars later evolving into classical WN stars as suggested by \cite{vink2023}, we may be looking at a snapshot of just such an evolutionary sequence.

We tested \texttt{BPASS} models of other metallicities within the confidence interval of the computed values. We found that the WR models favor lower metallicities, around 10\% $Z_{\odot}$, in which case a slightly larger age $\sim 5$ Myr is favored; and that the WR peaks are inconsistent with anything higher than 25\% $Z_{\odot}$.
We note that the observed WR bumps are brighter relative to the underlying continuum than the model spectrum from \texttt{BPASS}. This could in theory be due to stronger winds in the WR stars, but these winds are line driven and are weak at low metallicity. As is visible from \autoref{fig:bpass}, a binary population is needed to account for the observed WR emission strength, to compensate for the weaker wind stripping. 
\cite{adamo2024} show in their figure 2 an overview of inferred surface density and effective radius for a population of star clusters, including the Sunburst LCE. These authors find that the cluster is consistent with being a very massive proto-globular cluster with a high stellar density. This could statistically lead to an enhanced fraction of binary systems and in turn an enhanced VMS and/or WR population. Alternatively, a more top-heavy IMF than adopted in the fiducial \texttt{BPASS} models could also give rise to a larger population of WR stars.

We also note that the WR bumps of the Sunburst LCE are much broader than those of the model spectra. While the model spectra at these metallicities are clearly dominated by He\,\textsc{ii} and C\,\textsc{iv}, the observed bumps also show rich emission in the adjacent metal line complexes, in particular in N\,\textsc{iii} around 4650 Å. This shows significant nitrogen enhancement in the evolved massive stars of the cluster; consistent with the suggestion by \cite{kobayashi2024} that evolved, massive stars drive rapid nitrogen enhancement in the early phases of cluster evolution.

\subsection{The relative contributions from VMS and classic WR stars}
As previously stated, other authors have attributed the rest-frame UV stellar wind features to Very Massive Stars, based on the observation from low-redshift galaxies that the fraction of WR stars at low stellar metallicity is very low \citep[e.g.,][]{crowther2006,crowther2023}. \cite{mayya2023} have studied the star forming ring of the Cartwheel galaxy, which has a metallicity similar to that of the Sunburst LCE cluster. In this study, they calculated that the \ion{He}{ii} bump would have an $EW > 2$ Å only when a contingent of VMS was present. Our own comparison to the fiducial BPASS models in Sect.~\ref{sec:wrprops} suggests that a fraction of $\sim70\%$ WNh stars/VMS best simultaneously reproduces the combined Blue Bump equivalent width and bump intensity ratio, supporting that a large population of VMS is present. In contrast to this stands the morphology of the bumps, and especially the strength of the surrounding metal lines, primarily [\ion{N}{iii}]. The BPASS models fail to reproduce the bump morphology and the metal line strength at this metallicity, as is evident in \autoref{fig:bpass}. Also the \cite{martins2023} morphological classification scheme of the Blue and Orange bumps firmly classifies these bumps as stemming from classical WR stars. The \cite{martins2023} scheme is demonstrated and developed in the context of a galaxy of somewhat higher metallicity than the Sunburst LCE; however, the spectral features of the individual classes of VMS/WR stars are not expected to change with metallicity, only the overall and relative numbers of these star types. Given that the stellar wind features of VMS/WNh are a subset of those of classical WR stars, the \cite{martins2023} scheme cannot rule out a contribution from VMS; only suggest that it is considerably less dominant than in the BPASS models.

Additionally, regardless of the relative contributions from VMS and classic WR stars, it is clear that the fiducial BPASS model cannot reproduce the strength of the Blue and Orange bumps. It is necessary to invoke some mechanism or mechanisms which would give rise to a larger number of classic, hydrogen-stripped WR stars than this model; and in the previous section we suggested a higher binary fraction among the massive stars, due to the dense environment in this cluster. Such a boost to binary evolution could boost the population of classical WR stars considerably, and it could then lead one to ask whether invoking a VMS contribution would still be necessary to explain the bumps.

Clearly, closer analysis and higher signal-to-noise data would be necessary to properly settle this question; but tentatively, we can say with confidence that a surprisingly strong classic WR population is necessary to account for the observed bump morphologies, that a considerable contribution from VMS seems plausible but more difficult to ascertain, and that it seems likely that the 70\% VMS predicted by the fiducial BPASS model probably is an overestimate due to its failure to reproduce the broad metal emission features observed.


\subsection{Age from EW(H$\alpha$)}

These NIRSpec IFU spectra, covering the rest-frame optical in the Sunburst Arc, are the first detection of rest-frame optical continuum in this galaxy; allowing the first spectroscopic measurement of the equivalent width of H$\alpha$, \(EW_{0}(H\alpha) = 1040 \pm 31\) Å.
Given that the Sunburst LCE cluster is leaking ionizing radiation, this value should be corrected for
the amount of ionizing photons that have not been reabsorbed in the ISM giving rise to recombination lines. However, although the line-of-sight ionizing escape
fraction is high \citep{riverathorsen2019}; \citet{sunburst2017} have argued
based on the properties of Ly\(\alpha\) that the \emph{global} escape fraction is much lower
than that, perhaps as low as \(2\% \lesssim f_{\text{esc, glob}}^{\text{LyC}} \lesssim 5 \%\). 

Comparing to \texttt{STARBURST 99} models for the relevant metallicity \footnote{\url{https://www.stsci.edu/science/starburst99/figs/wha_inst_d.html}} shows
that the uncorrected EW(H\(\alpha\)) is consistent with a cluster age of \(\log(\text{age [yrs]})\sim 6.6\) or just under 4 Myr; within the same logarithmic age step best matching the WR emission. A more in-depth comparison of the age constraints imposed by H$\alpha$ and WR emission could potentially help constrain the global LyC escape fraction, and test the Ly$\alpha$ radiative transfer and ISM geometry scenario put forward by \cite{sunburst2017}.

\begin{table}[htbp]
\caption{\label{tab:derived}Derived physical properties of the Sunburst LCE cluster.}
\centering
\begin{tabular}{lc}
\toprule
Property & Value\\
\midrule
$z_{\text{LCE}}\tablefootmark{a}$ & $2.37108 \pm 2\times10^{-5}$\\
$E(B-V)_{\text{H}\alpha/\text{H}\beta}$ & $0.11 \pm 0.19$ \\
EW$_{0}$(H\(\alpha\)) [Å] & $1040 \pm 31$\\
Dynamical mass [10\textsuperscript{6} M\textsubscript{\(\odot\)}] & 9.0 \textpm{} 2.7\\
$T_e$ ([O\,\textsc{iii}]) [$10^4 K$] & $1.4 \pm 0.3$\\
$T_e$ ([N\,\textsc{ii}]) [$10^4 K$] & 1.34 \texttimes{} 10\textsuperscript{4}\\
$n_e$ ([O\,\textsc{ii}]) [cm$^{-2}$] & $1400^{+1500}_{-800}$\\
$n_e$ ([S\,\textsc{ii}]) [cm$^{-2}$] & $1000^{+600}_{-400}$\\
$P_{\text{ion}}/k_B$ [$10^7$ K/cm$^{3}$] & $3.6^{+2.6}_{-1.6}$\\
$12 + \log(\text{O/H})$ (O$^{+}+$ O$^{++}$) & $7.96^{+0.33}_{-0.21}$\\
$\log(\text{N/O})$ & -0.74 \textpm{} 0.09\\
\bottomrule
\end{tabular}
\tablefoot{
\tablefoottext{a}{Based on the full set of emission lines in G140H.}
}
\end{table}

\section{Summary and conclusion}
\label{sec:org9f2216e}

\begin{enumerate}
\item The Sunburst LCE cluster has a dynamical mass of $(9.0 \pm 1.1) \times 10^{6} M_{\odot}$. Accounting for typical mass loss over the age of local GCs \citep[e.g.,][]{adamo2024}, this brings the mass of the Sunburst LCE cluster very close to e.g., $\omega$ Cen.


\item The Sunburst LCE cluster has a moderately low oxygen abundance of $12 + \log(\text{O/H}) = 7.96^{+0.33}_{-0.21}$, or just under $20\%\, Z_{\odot}$, but an N/O ratio in the low-ionized medium slightly above solar.

\item The LCE cluster contains a population of massive Wolf-Rayet stars; to our
knowledge, this is the first time these have been directly and unambiguously identified using the rest-frame optical WR features at $z \gtrsim 0.5$. The relative strengths of the blue and the orange WR bumps, compared to
BPASS synthetic spectra, sets the LCE cluster age to slightly above 4~Myr.
Adopting the numbers from the BPASS model and scaling for cluster mass,
assuming that the dynamical mass is entirely stellar, yields a total of \(\sim\) 700 WR stars in the LCE cluster.

\item The BPASS model best fitting the blue-to-orange bump ratio contains as many as $\sim 70\%$ WNh stars, which are generally accepted as a spectroscopic designation of VMS. However, while these models can reproduce the relative bump strength well, they fail to reproduce the broad metal emission features surrounding the main \ion{He}{ii} and \ion{C}{iii} features. It is unclear whether the rapid nitrogen enrichment modeled by \cite{vink2023,kobayashi2024} can account for this discrepancy, but we tentatively suggest that the VMS/WR ratio predicted by the best BPASS model might be too high. 

\item The BPASS models with no binary evolution included produced a much poorer approximation to the observed spectra than those with binary evolution enabled. In particular, they all fail to produce a blue/orange ratio larger than 1.

\item The observed WR bump strength relative to the underlying continuum is stronger than in any of the model spectra, indicating that the Sunburst LCE cluster contains more, or more luminous, WR stars than assumed in the models. A number of possible explanations could conceivably account for this, e.g., a more top-heavy IMF, enhanced binary fraction, or stronger winds driven by rapid nitrogen enrichment. These data provide key observational constraints on stellar population models such as \texttt{BPASS}.

\item The LCE cluster is markedly nitrogen enriched for its low metallicity, with
log$(\text{N/O}) = -0.74\pm0.09$; about 0.8 dex higher than typical values for known
H\,\textsc{ii} regions of similar metallicity at low redshifts. The even
higher log$(\text{N/O})=-0.2$ in the dense, ionized condensations reported by \cite{pascale2023}
suggests that the majority of the nitrogen sits in compact, highly ionized
gas close to the most massive stars. This in turn lends credence to the
hypothesis that we are observing a rapid, ongoing nitrogen enrichment of the
cluster, driven mainly by very massive and bright stars, and lends credence to the hypothesis by \cite{ji2024} that most of the rapid nitrogen enrichment seen in galaxies such as gn-z11 is concentrated in high density and high-ionization regions around evolved high-mass stars, while the low-pressure and low-ionization evolves more moderately.

\end{enumerate}

These hints together show a picture of a proto-globular cluster at roughly 20\%
the current age of the Universe with a bright population of Wolf-Rayet stars, likely along with a significant contingent of Very Massive Stars, rapidly injecting nitrogen and other heavier elements into its ISM. The large WR star and VMS population indicates a generally large population of massive stars, possibly combined with a high binarity fraction, leading to a stronger and perhaps more protracted production of ionizing photons than accounted for by the models.

\begin{acknowledgements}
The authors want to thank A. Bik, A. Saldana-Lopez, and M. del Valle-Espinosa for helpful comments and suggestions. 
This work is  based on observations made with the  NASA/ESA/CSA James Webb Space
Telescope. The data were obtained from the Mikulski Archive for Space Telescopes
at the Space  Telescope Science Institute, which is operated  by the Association
of Universities for Research in Astronomy, Inc., under NASA contract NAS 5-03127
for JWST. These observations are associated with program GO 2555.
Support for program \#2555 was provided by NASA through a grant from the Space Telescope Science Institute, which is operated by the Association of Universities for Research in Astronomy, Inc., under NASA contract NAS 5-03127.
ER-T is supported by the Swedish Research Council grant No. 2022-04805\_VR.
\end{acknowledgements}

\bibliography{AllPapers}
\end{document}